# Coronary Artery Disease Research in India: A Scientometric Assessment of Publication during 1990-2019

Muneer Ahmad
muneerbangroo@gmail.com

Dr. M.Sadik Batcha
*Annamalai University*

# Coronary Artery Disease Research in India: A Scientometric Assessment of Publication during 1990-2019

**Muneer Ahmad[1] Dr. M Sadik Batcha[2]**

[1]*Research Scholar, Department of Library and Information Science, Annamalai University, Annamalai nagar, muneerbangroo@gmail.com*
[2]*Research Supervisor & Mentor, Professor and University Librarian, Annamalai University, Annamalai nagar, msbau@rediffmail.com*

**Abstract**

The present study examined 4698 Indian Coronary Artery Disease research publications, as indexed in Web of Science database during 1990-2019, with a view to understand their growth rate, global share, citation impact, international collaborative papers, distribution of publications by broad subjects, productivity and citation profile of top organizations and authors, and preferred media of communication. The Indian publications registered an annual average growth rate of 11.47%, global share of 1.14%, international collaborative publications share of 38.89% and its citation impact averaged to 25.58 citations per paper. Among broad subjects, Cardiovascular System & Cardiology contributed the largest publications share of 19.14% in Indian coronary artery disease output, followed by Neurosciences & Neurology (14.94%), Pharmacology & Pharmacy (8.51%), etc. during 1990-2019. Among various organizations and authors contributing to Indian coronary artery disease research, the top 20 organizations and top 30 authors together contributed 40.70% and 37.29% respectively as their share of Indian publication output and 38.36% and 33.13% respectively as their share of Indian citation output during 1990-2019. Among 1222 contributing journals in Indian coronary artery disease research, the top 30 journals registered 30.80% share during 1990-2019. There is an urgent need to increase the publication output, improve research quality and improve international collaboration. Indian government also needs to come up with a policy for identification, screening, diagnosis and treatment of coronary artery disease patients, besides curriculum reform in teaching, capacity building, patient education and political support are badly needed.

**Introduction**

*Coronary artery disease (CAD)*-often called coronary heart disease or CHD, is generally used to refer to the pathologic process affecting the coronary arteries (usually atherosclerosis). CAD includes the diagnoses of angina pectoris, myocardial infarction (MI), silent myocardial ischemia, and CAD mortality that result from CAD. Hard CAD endpoints generally include MI and CAD death. The term CHD is often used interchangeably with CAD. *CAD death*—Includes sudden cardiac death (SCD) for circumstances when the death has occurred within 24 hours of the abrupt onset of symptoms, and the term non- SCD applies when the time course from the clinical presentation until the time of death exceeds 24 hours or has not been specifically identified. *Atherosclerotic cardiovascular disease (ASCVD, often shortened to CVD)*-the pathologic process affecting the entire arterial circulation, not just the coronary arteries. Stroke, transient ischemic attacks, angina, MI, CAD death, claudication, and critical limb ischemia are manifestations of ASCVD (Lemos & Omland, 2018).

Coronary artery disease (CAD) is a major cause of death and disability in developed countries. Although CAD mortality rates worldwide have declined over the past 4 decades, CAD remains responsible for approximately one-third or more of all deaths in individuals over age 35, and it has been estimated that nearly half of all middle-aged men and one-third of middle aged women in the United States will develop clinical CAD (Mozaffarian et al., 2016). A Global Burden of Disease Study Group report from 2013 estimated that 17.3 million deaths worldwide in 2013 were related to ASCVD, a 41% increase since 1990 (GBD: 2013 Mortality and Causes of Death Collaborators, 2015). Although the absolute numbers of ASCVD deaths had increased significantly since 1990, the age-standardized death rate decreased by 22% in the same period, primarily due to shifting age demographics and causes of death worldwide (Towfighi, Zheng, & Ovbiagele, 2009).

Heart disease mortality has declined since the 1970s in the United States and in regions where economies and healthcare systems are relatively advanced. Ischemic heart disease remains the number one cause of death in adults on a worldwide basis (GBD: 2013 Mortality and Causes of Death Collaborators, 2015). In a 2014 study using World Health Organization data from 49 countries in Europe and northern Asia, over 4 million annual deaths were attributable to ASCVD (Nichols, Townsend, Scarborough, & Rayner, 2014). Current worldwide estimates for heart disease mortality show Eastern European countries have the highest ASCVD death rates (> 200 per 100,000/year), followed by an intermediate group that includes most countries with modern economies (100–200 per 100,000/ year), and the lowest levels (0–100 per 100,000/year) are largely observed in European countries and a few non- European countries with advanced healthcare systems. A detailed analysis of European country specific data showed that CHD mortality rates dropped by more than 50% over the 1980–2009 interval, and the decline was observed across virtually all European countries for both sexes. The authors of the report concluded that the downward trends did not appear to show a plateau. Rather, CHD mortality was stable or continuing to decline across Europe (Nichols, Townsend, Scarborough, & Rayner, 2013). Complementary analyses have been undertaken in the United States, and CHD mortality has been demonstrated to have peaked in the 1970s and declined since that date (Mozaffarian et al., 2016).

**Indian Perspective**

The office of the RGI has periodically reported data on cardiovascular mortality rates in India (Registrar General of India, 2013). These data have been summarized as circulatory system deaths in the Medical Certification of Cause of Deaths reports, and in 1980s and 1990s it was reported that CVD led to 15%-20% of deaths in the country (Gupta, Misra, Pais, Rastogi, & Gupta, 2006). An increasing trend in proportionate CVD mortality has been reported, with 20.6% deaths in 1990, 21.4% in 1995, 24.3% in 2000, 27.5% in 2005, and 29.0% in 2013 (Registrar General of India, 2013).

However, these reports were based on incomplete data (mainly rural health surveys) from which national data were extrapolated. The Million Death Study Group in collaboration with RGI reported deaths for the year 2001-2003 using a validated verbal autopsy instrument (Registrar General of India, 2013). This study used the existing sample registration surveys of the Indian government and evaluated more than 120,000 death reports obtained from 661 districts of the country using a nationally representative sample of more than 6 million participants. CVD emerged as the most important cause of death in men and women, in urban and rural populations, and in developed and developing states of the country (Registrar General of India, 2013). In India, more than 10.5 million deaths occur annually, and it was reported that CVD led to 20.3% of these deaths in men and 16.9% of all deaths in women (Registrar General of India, 2013)**.** According to 2010-2013 RGI data, (Registrar General of India, 2011) proportionate mortality from CVD increased to 23% of total and 32% of adult deaths in years 2010-2013. The mortality varies from <10% in rural locations in less developed states to >35% in more developed urban locations(Institute for Health Metrics and Evaluation (IHME), 2014). Geographic distribution of CVD mortality in India indicates that in less developed regions, such as the eastern and northeastern states with low Human Development indices, there is lower proportionate mortality compared with better developed states in southern and western regions. There is a linear relationship of increasing proportionate CVD mortality with regional Human Development Index, which confirms the presence of the epidemiological transition introduced earlier (Gaziano & Gaziano, 2008; Kuate Defo, 2014).

**Literature Review**

The review, in general, provides an overview of the theory and the research literature, with a special emphasis on the literature specific to the topic of investigation. It provides support to the proposition of one's research, with ample evidences drawn from subject experts and authorities in the concerned field. The sources consulted for the review of literature here includes Scientometric studies related materials drawn from Primary periodicals.

(Batcha & Ahmad, 2017) obtained the analysis of two journals Indian Journal of Information Sources and Services (IJSS) which is of Indian origin and Pakistan Journal of Library and Information Science (PJLIS) from Pakistan origin and studied them comparatively with scientometric indicators like year wise distribution of articles, pattern of authorship and productivity, degree of collaboration, pattern of co-authorship, average length of papers, average keywords, etc and found 138 (94.52%) of contributions from IJISS were made by Indian authors and similarly 94 (77.05) of contributions from PJLIS were done by Pakistani authors. The collaboration with foreign authors of both the countries is negligible (1.37% of articles) from India and (4.10% of articles) from Pakistan.

(Ahmad, Batcha, Wani, Khan, & Jahina, 2018) studied Webology journal one of the reputed journals from Iran through scientometric analysis. The study aims to provide a comprehensive analysis regarding the journal like year wise growth of research articles, authorship pattern, author productivity, and subjects taken by the authors over the period of 5 years from 2013 to 2017. The findings indicate that 62 papers were published in the journal during the study period. The articles having collaborative nature were high in number. Regarding the subject concentration of papers of the journal, Social Networking, Web 2.0, Library 2.0 and Scientometrics or Bibliometrics were highly noted. The results were formulated through standard formulas and statistical tools.

(Batcha, Jahina, & Ahmad, 2018) has examined the DESIDOC Journal by means of various scientometric indicators like year wise growth of research papers , authorship pattern, subjects and themes of the articles over the period of five years from 2013 to 2017. The study reveals that 227 articles were published over the five years from 2013 to 2017. The authorship pattern was highly collaborative in nature. The maximum numbers of articles (65 %) have ranged their thought contents between 6 and 10 pages.

(Ahmad & Batcha, 2019) analyzed research productivity in Journal of Documentation (JDoc) for a period of 30 years between 1989 and 2018. Web of Science a service from Clarivate Analytics has been consulted to obtain bibliographical data and it has been analysed through Bibexcel and Histcite tools to present the datasets. Analysis part deals with local and global citation level impact, highly prolific authors and their research output, ranking of prominent institution and countries. In addition to this scientographical mapping of bibliographical data is obtainable through VOSviewer, which is open source mapping software.

(Ahmad & Batcha, 2019) studied the scholarly communication of Bharathiar University which is one of the vibrant universities in Tamil Nadu. The study find out the impact of research produced, year-wise research output, citation impact at local and global level, prominent authors and their total output, top journals of publications, top collaborating countries which collaborate with the university authors, highly industrious departments and trends in publication of the university during 2009 through 2018. During the 10 years of study under consideration it indicates that a total of 3440 research articles have been published receiving 38104 citations having h-index as 68. In addition the study used scientographical mapping of data and presented it through graphs using VOSviewer software mapping technique.

(Ahmad, Batcha, & Jahina, 2019) quantitatively measured the research productivity in the area of artificial intelligence at global level over the study period of ten years (2008-2017). The study acknowledged the trends and features of growth and collaboration pattern of artificial intelligence research output. Average growth rate of artificial intelligence per year increases at the rate of 0.862. The multi-authorship pattern in the study is found high and the average number of authors per paper is 3.31. Collaborative Index is noted to be the highest range in the year 2014 with 3.50. Mean CI during the period of study is 3.24. This is also supported by the mean degree of collaboration at the percentage of 0.83 .The mean CC observed is 0.4635. Regarding the application of Lotka's Law of authorship productivity in the artificial intelligence literature it proved to be fit for the study. The distribution frequency of the authorship follows the exact Lotka's Inverse Law with the exponent á = 2. The modified form of the inverse square law, i.e.,

Inverse Power Law with á and C parameters as 2.84 and 0.8083 for artificial intelligence literature is applicable and appears to provide a good fit. Relative Growth Rate [Rt(P)] of an article gradually increases from -0.0002 to 1.5405, correspondingly the value of doubling time of the articles Dt(P) decreases from 1.0998 to 0.4499 (2008-2017). At the outset the study reveals the fact that the artificial intelligence literature research study is one of the emerging and blooming fields in the domain of information sciences.

(Batcha, Dar, & Ahmad, 2019) presented a scientometric analysis of the journal titled "Cognition" for a period of 20 years from 1999 to 2018. The study was conducted with an aim to provide a summary of research activity in the journal and characterize its most aspects. The research coverage includes the year wise distribution of articles, authors, institutions, countries and citation analysis of the journal. The analysis showed that 2870 papers were published in journal of Cognition from 1999 to 2018. The study identified top 20 prolific authors, institutions and countries of the journal. Researchers from USA have made the most percentage of contributions.

**Objectives**

The present manuscript aims to study the various dimensions of Indian coronary artery disease research output in terms of various bibliometric indicators, based on publications and citation data, derived from Web of Science database during 1990-2020. In particular, the study analyzed overall annual and cumulative growth of Indian publications, its global share among top 6 most productive countries, its citation impact, its international collaborative papers share, publication output distribution by broad sub-fields, productivity and citation impact of most productive organizations and authors, and leading media of communications.

**Methodology**

For the present study, the publication data was retrieved and downloaded from the Web of Science database (http://apps.webofknowledge.com/) on coronary artery disease research during 1990-2020. A main search strategy for global output was formulated, where the keyword such as ("coronary artery disease'') and mesh terms ("coronary arteriosclerosis" OR "coronary atherosclerosis" OR "coronary ischemic" OR ''arterial sclerosis'' AND CU="India") were searched together in the "Topic tag" and further limited the search output to period '1990-2019' within "date range tag". This search strategy generated 4698 Indian publications on coronary artery disease from the Web of Science database. Detailed analysis was carried out on 4698 Indian publications using the Histcite and Bibexcel tools to get data distribution by subject, collaborating countries, author-wise, organization-wise and journal-wise, etc. Further, mapping tool such as VOSviewer was used to study the collaboration behavior and citation network.

**Analysis**

The global and Indian research output in coronary artery disease research cumulated to 411668 and 4698 publications in 30 years during 1990-2019 and they increased from 1801 and 15 in the year 1990 to 22483 and 390 publications in the year 2019, registering 8.77% and 11.47% growth per annum. Their ten-year cumulative output increased from 69679 and 331 to 133574 and 1136 to 208415 and 3231 publications from 1990-1999 to 2000-2009 to 2010-2019, registering 6.72%, 13.12%, 4.55% and 11.02% growth respectively. The share of Indian publications in global output was 1.14% during 1990-2019, which increased from 0.48% to 0.85% to 1.55% from 1990-1999 to 2000-2009 to 2010-2019 respectively. Amongst Indian publications on coronary artery disease, 69.9% (3286) was published as articles, 11.7% (551) as meeting abstract, 11.4% (535) as review, 2.8% (133) as letter, 2.0% (94) as editorial material, 1.3% (59) as article; proceedings paper, 0.2%(11) as article; early access, 0.2% (8) as note, 0.1% (5) review; early access, 0.1% (4) Article; retracted publication and correction, 0.1% (3) Review; book chapter, 0.0% (2) article; book chapter and reprint and 0.0% (1) biographical-item . The research impact as measured by citations per paper registered by Indian publications in coronary artery disease averaged to 25.58 citations per publication (CPP) during 1990-2019; ten-yearly impact averaged to 21.46 CPP for the period 1990-1999 which increased to 30.38 CPP in the succeeding ten-year 2000-2009 and then declined to 24.32 CPP for the period 2010-2019 (Table 1).

*Table 1: World and India's Output in Coronary Artery Disease Research, 1990-2019.*

| Publication Period | World | India | | | |
|---|---|---|---|---|---|
| | TP | TP | TGCS | CPP | %TP |
| 1990 | 1801 | 15 | 80 | 5.33 | 0.83 |
| 1991 | 5434 | 24 | 303 | 12.63 | 0.44 |
| 1992 | 5929 | 21 | 270 | 12.86 | 0.35 |
| 1993 | 6236 | 25 | 361 | 14.44 | 0.40 |
| 1994 | 6756 | 26 | 484 | 18.62 | 0.38 |
| 1995 | 7255 | 24 | 821 | 34.21 | 0.33 |
| 1996 | 7804 | 45 | 1137 | 25.27 | 0.58 |
| 1997 | 9154 | 49 | 1421 | 29.00 | 0.54 |

| Year | TP | TC | ICP | CPP | |
|---|---|---|---|---|---|
| 1998 | 9384 | 53 | 1393 | 26.28 | 0.56 |
| 1999 | 9926 | 49 | 834 | 17.02 | 0.49 |
| 2000 | 10656 | 55 | 1543 | 28.05 | 0.52 |
| 2001 | 10564 | 57 | 1792 | 31.44 | 0.54 |
| 2002 | 10771 | 60 | 2080 | 34.67 | 0.56 |
| 2003 | 11837 | 72 | 1299 | 18.04 | 0.61 |
| 2004 | 12977 | 101 | 2580 | 25.54 | 0.78 |
| 2005 | 13662 | 110 | 2260 | 20.55 | 0.81 |
| 2006 | 14362 | 103 | 4441 | 43.12 | 0.72 |
| 2007 | 15130 | 143 | 5202 | 36.38 | 0.95 |
| 2008 | 16241 | 212 | 7381 | 34.82 | 1.31 |
| 2009 | 17374 | 223 | 5932 | 26.60 | 1.28 |
| 2010 | 17695 | 255 | 6281 | 24.63 | 1.44 |
| 2011 | 18664 | 265 | 11131 | 42.00 | 1.42 |
| 2012 | 19071 | 307 | 17519 | 57.07 | 1.61 |
| 2013 | 21172 | 314 | 14773 | 47.05 | 1.48 |
| 2014 | 20683 | 330 | 5355 | 16.23 | 1.60 |
| 2015 | 21689 | 272 | 5199 | 19.11 | 1.25 |
| 2016 | 22363 | 421 | 7456 | 17.71 | 1.88 |
| 2017 | 22420 | 312 | 7797 | 24.99 | 1.39 |
| 2018 | 22175 | 365 | 2462 | 6.75 | 1.65 |
| 2019 | 22483 | 390 | 611 | 1.57 | 1.73 |
| 1990-1999 | 69679 | 331 | 7104 | 21.46 | 0.48 |
| 2000-2009 | 133574 | 1136 | 34510 | 30.38 | 0.85 |
| 2010-2019 | 208415 | 3231 | 78584 | 24.32 | 1.55 |
| 1990-2019 | 411668 | 4698 | 120198 | 25.58 | 1.14 |

*TP: Total Papers; TC: Total Citations; CPP: Citations Per Paper; ICP: International Collaborative Papers*

**Publication Profile of Top 6 Most Productive Countries**

More than 140 countries of the world participated in global research in coronary artery disease research during 1990-2019. Between 4698 and 139222 publications were contributed by top 6 most productive countries in coronary artery disease research and they together accounted for 65.69% of global publication share during 1990-2019. Their ten-year publications output decreased from 65.34% to 63.27% from 1990-1999 to 2000-2009 and then increased 67.35% in 2010-2019. Each of top 6 countries had global publication share between 1.14% and 33.82% during 1990-2019. USA accounted for the highest publication share (33.82%), followed by Germany (8.12%), Republic of China (8.02%), Japan (7.61%), England (6.97%) and India

(1.14%) during 1990-2019. Their ten-year global publication share have increased by 2.45% in Republic of China, followed by India (0.38%), Germany (0.28%), and England (0.04%), as against decline by 4.83% in USA and 0.37% in Japan from 1990-1999 to 2000-2009 and then again ten-year global share have increased by 10.61% in Republic of China, followed by India (0.70%) and England (0.01%) as against decline by 4.27% in USA, 1.54% in Japan and 1.44% in Germany from 2000-2009 to 2010-2019 (Table 2).

*Table 2: Global Publication Output and Share of Top 6 Countries in Coronary Artery Disease Research during 1990-2019*

| S.No. | Country Name | TP | | | | %TP | | | |
|---|---|---|---|---|---|---|---|---|---|
| | | 1990-1999 | 2000-2009 | 2010-2019 | 1990-2019 | 1990-1999 | 2000-2009 | 2010-2019 | 1990-2019 |
| 1 | USA | 27869 | 46971 | 64382 | 139222 | 40.00 | 35.16 | 30.89 | 33.82 |
| 2 | Germany | 6008 | 11886 | 15553 | 33447 | 8.62 | 8.90 | 7.46 | 8.12 |
| 3 | Republic of China | 430 | 4093 | 28506 | 33029 | 0.62 | 3.06 | 13.68 | 8.02 |
| 4 | Japan | 6063 | 11126 | 14158 | 31347 | 8.70 | 8.33 | 6.79 | 7.61 |
| 5 | England | 4825 | 9305 | 14547 | 28677 | 6.92 | 6.97 | 6.98 | 6.97 |
| 6 | India | 331 | 1136 | 3231 | 4698 | 0.48 | 0.85 | 1.55 | 1.14 |
| | Total of 6 Countries | 45526 | 84517 | 140377 | 270420 | 65.34 | 63.27 | 67.35 | 65.69 |
| | World Output | 69679 | 133574 | 208415 | 411668 | 100.00 | 100.00 | 100.00 | 100.00 |
| | Share of 6 in World Output | 65.34 | 63.27 | 67.35 | | 65.34 | 63.27 | 67.35 | |

**India's International Collaboration**

The share of India's international collaborative publications (ICP) in its national output in coronary artery disease research was 38.88% during 1990-2019, which increased from 0.89% during 1990-1999 to 6.00% during 2000-2009 and then again increased to 31.99% during 2010-2019. About 139 foreign countries collaborated with India in 1827 coronary artery disease research papers during 1990-2019. These 1827 papers together registered 285,336 citations, with 156 citations per paper. USA, among foreign countries, contributed the largest share (40.45%) to India's international collaborative papers in coronary artery disease research, followed by England (14.72%), Canada (14.07%), Peoples Republic of China (10.73%), Australia (10.24%), and Germany (9.80%) during 1990-2019. The share of ICP increased by 7.40% in Canada, followed by 5.57% in USA, 4.91% in England, as against decrease by 10.03% Republic of China, 4.46% in Australia and 3.39% in Germany from 1990-1999 to 2000-2009 and then again share of ICP increased by 7.59% in Peoples Republic of China, followed by 3.38% in England,

3.27% in Australia and 1.47% in Germany, as against decrease by 15.33% USA and 0.37% in Canada from 2000-2009 to 2010-2019 (Table 3).

*Table 3: The Share of Top 6 Foreign Countries in India's International Collaborative Papers in India's Coronary Artery Disease Research during 1990-2019.*

| S.No. | Collaborative Country | Number of International Collaborative Papers | | | | Share of International Collaborative Papers | | | |
|---|---|---|---|---|---|---|---|---|---|
| | | 1990-1999 | 2000-2009 | 2010-2019 | 1990-2019 | 1990-1999 | 2000-2009 | 2010-2019 | 1990-2019 |
| 1 | USA | 20 | 150 | 569 | 739 | 47.62 | 53.19 | 37.86 | 40.45 |
| 2 | England | 3 | 34 | 232 | 269 | 7.14 | 12.06 | 15.44 | 14.72 |
| 3 | Canada | 3 | 41 | 213 | 257 | 7.14 | 14.54 | 14.17 | 14.07 |
| 4 | Peoples Republic of China | 6 | 12 | 178 | 196 | 14.29 | 4.26 | 11.84 | 10.73 |
| 5 | Australia | 5 | 21 | 161 | 187 | 11.90 | 7.45 | 10.71 | 10.24 |
| 6 | Germany | 5 | 24 | 150 | 179 | 11.90 | 8.51 | 9.98 | 9.80 |
| | Total | 42 | 282 | 1503 | 1827 | 100.00 | 100.00 | 100.00 | 100.00 |

**Subject-Wise Distribution of Indian Research Output**

As per the Web of Science database classification, India's coronary artery disease research output is distributed across 88 subjects during 1990-2019. Among subjects, cardiovascular system and cardiology registered the highest publications share (19.14%), followed by neurosciences and neurology (14.94%), pharmacology and pharmacy (8.51%), general and internal medicine (4.40%), biochemistry and molecular biology (4.22%), research and experimental medicine (3.81%), surgery (3.23%), cell biology (2.96%), endocrinology and metabolism (2.76%), pediatrics (2.12%) and other subjects respectively during 1990-2019 (Table 4).

*Table 4: Subject-Wise Breakup of Indian Publications in Coronary artery Disease Research during 1990-2019*

| S.No. | *Subject wise | TP | % |
|---|---|---|---|
| 1 | Cardiovascular System & Cardiology | 1298 | 19.14 |
| 2 | Neurosciences & Neurology | 1013 | 14.94 |
| 3 | Pharmacology & Pharmacy | 577 | 8.51 |
| 4 | General & Internal Medicine | 298 | 4.40 |
| 5 | Biochemistry & Molecular Biology | 286 | 4.22 |
| 6 | Research & Experimental Medicine | 258 | 3.81 |
| 7 | Surgery | 219 | 3.23 |
| 8 | Cell Biology | 201 | 2.96 |
| 9 | Endocrinology & Metabolism | 187 | 2.76 |
| 10 | Pediatrics | 144 | 2.12 |

| 11 | Science & Technology - Other Topics | 125 | 1.84 |
| 12 | Engineering | 121 | 1.78 |
| 13 | Immunology | 120 | 1.77 |
| 14 | Hematology | 119 | 1.76 |
| 15 | Genetics & Heredity | 111 | 1.64 |
| 16 | Radiology, Nuclear Medicine & Medical Imaging | 98 | 1.45 |
| 17 | Nutrition & Dietetics | 86 | 1.27 |
| 18 | Respiratory System | 85 | 1.25 |
| 19 | Chemistry | 71 | 1.05 |
| 20 | Ophthalmology | 69 | 1.02 |

*There is overlapping of literature covered under various subjects

**Significant Keywords**

Around 7357 significant keywords have been identified from the literature, which highlight possible research trends in Indian coronary artery disease research. The 40 keywords are listed in table 5 in the decreasing order of their frequency of occurrence in 30 years during 1990-2019.

*Table 5: List of Significant Keywords in Literature on Indian Coronary Artery Disease Research during 1990-2019.*

| S.No. | Name of Key Words | Frequency | S.No. | Name of Key Words | Frequency |
|---|---|---|---|---|---|
| 1 | Coronary | 981 | 21 | Cerebral | 214 |
| 2 | Disease | 956 | 22 | Population | 214 |
| 3 | Ischemic | 936 | 23 | Cardiovascular | 203 |
| 4 | Stroke | 857 | 24 | Rats | 191 |
| 5 | Artery | 850 | 25 | Factors | 176 |
| 6 | Patients | 719 | 26 | Reperfusion | 173 |
| 7 | Acute | 437 | 27 | Clinical | 168 |
| 8 | Risk | 409 | 28 | Diabetes | 167 |
| 9 | Indian | 359 | 29 | Infarction | 163 |
| 10 | Effect | 288 | 30 | Case | 147 |
| 11 | India | 277 | 31 | North | 142 |
| 12 | Myocardial | 276 | 32 | Polymorphism | 142 |
| 13 | Heart | 263 | 33 | Using | 142 |
| 14 | Association | 249 | 34 | Based | 141 |
| 15 | Gene | 236 | 35 | South | 138 |
| 16 | Ischemia | 228 | 36 | Cardiac | 137 |
| 17 | Induced | 224 | 37 | Stress | 137 |
| 18 | Role | 223 | 38 | Therapy | 135 |
| 19 | Injury | 215 | 39 | Trial | 135 |
| 20 | Analysis | 214 | 40 | Rat | 132 |

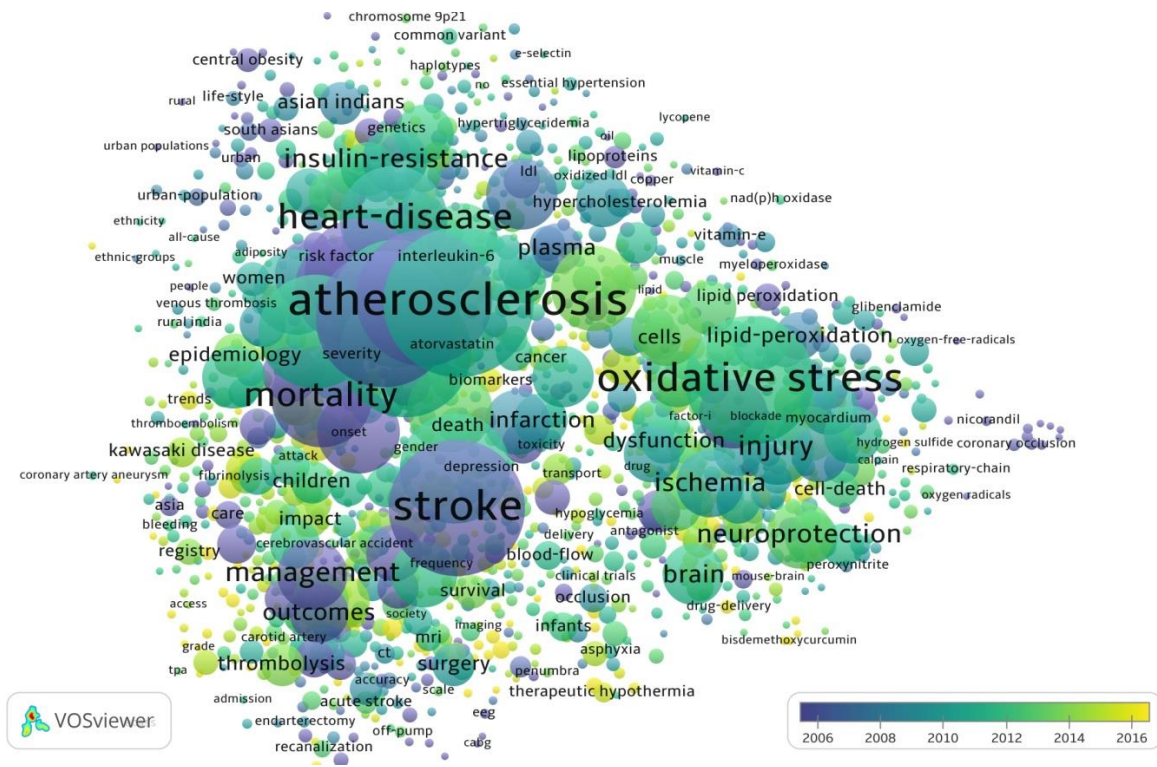

*Cluster of Keywords on Indian Coronary Artery Disease Research*

**Profile of Top 20 Most Productive Indian Organizations**

The top 20 Indian organizations contribution to coronary artery disease research varied from 45 to 496 publications and they together accounted for 40.10% (1887) publication share and 82.20% (98807) citation share to its cumulative publications output during 1990-2019. Table 6 presents a scientometric profile of these 20 India organizations.

*Table 6: Scientometric Profile of Top 20 Most Productive Indian Organizations in Coronary Artery Disease Research during 1990-2019*

| S.No. | Name of Organization | TP | % | TGCS | CPP |
|---|---|---|---|---|---|
| 1 | All India Institute of Medical Sciences (AIIMS), New Delhi | 496 | 10.60 | 29360 | 59.19 |
| 2 | Postgraduate Institute of Medical Education & Research (PGIMER), Chandigarh | 212 | 4.50 | 8165 | 38.51 |
| 3 | Christian Medical College & Hospital, Vellore | 121 | 2.60 | 8814 | 72.84 |
| 4 | Sree Chitra Tirunal Institute Medical Science & Technology | 105 | 2.20 | 6108 | 58.17 |
| 5 | Nizams Institute of Medical Science | 104 | 2.20 | 1727 | 16.61 |
| 6 | Sanjay Gandhi Postgraduate Institute Medical Science | 100 | 2.10 | 18847 | 188.47 |
| 7 | Osmania University | 67 | 1.40 | 866 | 12.93 |
| 8 | Natl Institution Mental Health & NeuroScience | 62 | 1.30 | 1046 | 16.87 |

| 9 | Madras Diabet Research Foundation | 57 | 1.20 | 5028 | 88.21 |
|---|---|---|---|---|---|
| 10 | Banaras Hindu University | 56 | 1.20 | 3866 | 69.04 |
| 11 | Panjab University | 56 | 1.20 | 1478 | 26.39 |
| 12 | University Delhi | 53 | 1.10 | 4382 | 82.68 |
| 13 | GB Pant Hospital | 52 | 1.10 | 628 | 12.08 |
| 14 | Manipal University | 52 | 1.10 | 902 | 17.35 |
| 15 | Govt Medical College | 51 | 1.10 | 1514 | 29.69 |
| 16 | Post Graduate Institution of Medical Education & Research | 51 | 1.10 | 3057 | 59.94 |
| 17 | Punjabi University | 50 | 1.10 | 629 | 12.58 |
| 18 | Sir Ganga Ram Hospital | 49 | 1.00 | 694 | 14.16 |
| 19 | Apollo Hospital | 48 | 1.00 | 813 | 16.94 |
| 20 | Maulana Azad Medical College | 45 | 1.00 | 883 | 19.62 |
| | Total of 20 Organizations | 1887 | 40.10 | 98807 | 52.36 |
| | Total of India | 4698 | 100 | 270420 | 57.56 |
| | Share of 20 Organizations in Indian total output | 40.17 | | 36.54 | |

*TP: Total Papers; TGCS: Total Global Citations Score; CPP: Citations Per Paper*

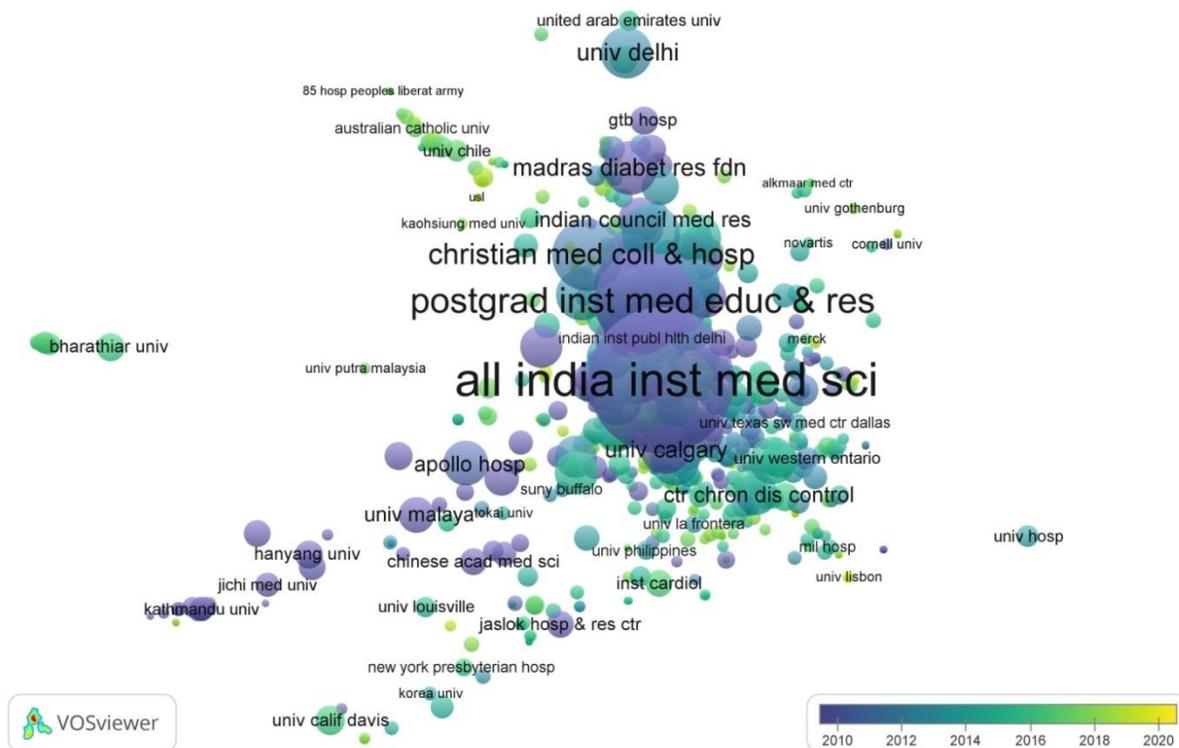

*Cluster of Most Productive Indian Organizations in Coronary Artery Disease Research*

Six organizations registered higher productivity than the group average of 94.35: All India Institute of Medical Sciences (AIIMS), New Delhi (496 papers), Postgraduate Institute of

Medical Education & Research (PGIMER), Chandigarh (212 papers), Christian Medical College & Hospital, Vellore (121 papers), Sree Chitra Tirunal Institute Medical Science & Technology (105 papers), Nizams Institute of Medical Science (104 papers), Sanjay Gandhi Postgraduate Institute Medical Science (100) during 1990-2019.

Eight organizations registered higher citation impact than the group average of 52.36: All India Institute of Medical Sciences (AIIMS), New Delhi (59.19), Christian Medical College & Hospital, Vellore (72.84), Sree Chitra Tirunal Institute Medical Science & Technology (58.17), Sanjay Gandhi Postgraduate Institute Medical Science (188.47), Madras Diabet Research Foundation (88.21), Banaras Hindu University (69.04), University Delhi (82.68) and Post Graduate Institution of Medical Education & Research (59.94) during 1990-2019.

**Profile of Top 30 Most Productive Authors**

The top 30 Indian author's contribution to coronary artery disease research varied from 35 to 146 publications and they together accounted for 37.29% (1752) publication share and 38.36% (103744) citation share to its cumulative publications output during 1990-2019. Table 7 presents a scientometric profile of these 20 India authors.

*Table 7: Scientometric Profile of Top 20 Most Productive Authors in Coronary Artery Disease Research during 1990-2019*

| S.No. | Authors | TP | % | TGCS | CPP |
|---|---|---|---|---|---|
| 1 | Kumar A | 146 | 3.1 | 2233 | 15.29 |
| 2 | Kumar S | 100 | 2.1 | 732 | 7.32 |
| 3 | Prasad K | 87 | 1.9 | 1090 | 12.53 |
| 4 | Singh S | 80 | 1.7 | 1197 | 14.96 |
| 5 | Sharma A | 78 | 1.7 | 590 | 7.56 |
| 6 | Kaul S | 74 | 1.6 | 958 | 12.95 |
| 7 | Mohan V | 74 | 1.6 | 6091 | 82.31 |
| 8 | Kumar P | 70 | 1.5 | 2881 | 41.16 |
| 9 | Singh N | 68 | 1.4 | 3187 | 46.87 |
| 10 | Singh RB | 68 | 1.4 | 2611 | 38.40 |
| 11 | Singh M | 63 | 1.3 | 1182 | 18.76 |
| 12 | Gupta R | 62 | 1.3 | 20427 | 329.47 |
| 13 | Gupta A | 61 | 1.3 | 713 | 11.69 |
| 14 | Prabhakaran D | 59 | 1.3 | 8343 | 141.41 |
| 15 | Sylaja PN | 53 | 1.1 | 1390 | 26.23 |
| 16 | Bhatia R | 51 | 1.1 | 875 | 17.16 |
| 17 | Das S | 50 | 1.1 | 993 | 19.86 |
| 18 | Sharma S | 47 | 1 | 282 | 6.00 |
| 19 | Pandian JD | 44 | 0.9 | 16111 | 366.16 |
| 20 | Xavier D | 43 | 0.9 | 9570 | 222.56 |
| 21 | Munshi A | 40 | 0.9 | 670 | 16.75 |
| 22 | Karthikeyan G | 39 | 0.8 | 14634 | 375.23 |
| 23 | Gupta S | 38 | 0.8 | 801 | 21.08 |
| 24 | Jaggi AS | 38 | 0.8 | 592 | 15.58 |
| 25 | Kapoor A | 38 | 0.8 | 564 | 14.84 |
| 26 | Trehan N | 38 | 0.8 | 901 | 23.71 |
| 27 | Khurana D | 36 | 0.8 | 131 | 3.64 |
| 28 | Kumar R | 36 | 0.8 | 989 | 27.47 |
| 29 | Niaz MA | 36 | 0.8 | 1849 | 51.36 |
| 30 | Ghosh S | 35 | 0.7 | 1157 | 33.06 |
| Total of 30 authors | | 1752 | 37.29 | 103744 | 59.21 |
| Total of India | | 4698 | | 270420 | 57.56 |
| Share of 30 authors in India's output | | 37.29 | | 38.36 | |

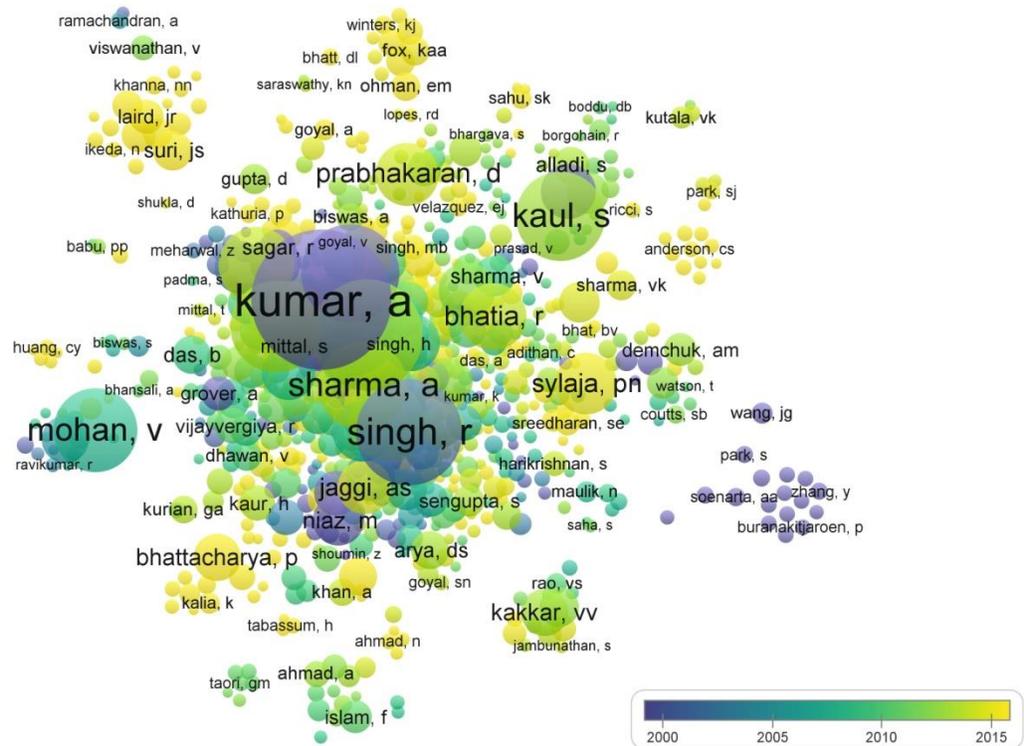

*Cluster of Most Productive Authors in Coronary Artery Disease Research*

Fourteen authors registered higher publications productivity than group average of 58.4: Singh S (80 papers), Sharma A (78 papers), Kaul S and Mohan V (74 papers each), Kumar P (70 papers), Singh N and Singh RB (68 papers each), Singh M (63 papers), Gupta R (62 papers), Gupta A (61 papers), and Prabhakaran D (59 papers) during 1990-2019.

Six authors registered higher citation impact than the group average of 59.21 citations per publication: Karthikeyan G (375.23), Pandian D (366.16), Gupta R (329.47), Xavier D (222.56), Prabhakaran D (141.41) and Mohan V (82.31) during 1990-2019.

**Medium of Communication**

Among India's coronary artery disease output, Indian publications on coronary artery disease, 69.9% (3286) was published as articles, 11.7% (551) as meeting abstract, 11.4% (535) as review, 2.8% (133) as letter, 2.0% (94) as editorial material, 1.3% (59) as article; proceedings paper, and other forms are less than one percent. The top 30 most productive journals accounted for 22 to 143 papers. The top 30 journals publishing Indian papers in coronary artery disease together accounted for 30.80% share (1447 papers) of total Indian journal publication output during 1990-2019. Neurology India *was* the most productive journals with 143 papers, followed by International Journal of Cardiology (139 papers), Annals of Indian Academy of Neurology (99 papers), Journal of the Neurological Sciences (77 papers), Indian Journal of Medical Research (69 papers), Molecular And Cellular Biochemistry (56 papers) Stroke (55 papers), Journal of the American College of Cardiology and PLOS One (52 papers each) etc. during 1990-2019 (Table 8).

*Table 8: Productivity of Top 30 Most Productive Journals in Indian Coronary Artery Disease Research during 1990-2019*

| S.No. | Name of the Journals | Number of Papers | % | TLCS | TGCS |
|---|---|---|---|---|---|
| 1 | Neurology India | 143 | 3.00 | 135 | 945 |
| 2 | International Journal of Stroke | 139 | 3.00 | 52 | 682 |
| 3 | International Journal of Cardiology | 108 | 2.30 | 131 | 2110 |
| 4 | Annals of Indian Academy of Neurology | 99 | 2.10 | 40 | 559 |
| 5 | Journal of The Neurological Sciences | 77 | 1.60 | 105 | 573 |

| S.No | Journal | TP | % | TPC | TC |
|---|---|---|---|---|---|
| 6 | Indian Journal of Medical Research | 69 | 1.50 | 87 | 1393 |
| 7 | Molecular And Cellular Biochemistry | 56 | 1.20 | 71 | 1201 |
| 8 | Stroke | 55 | 1.20 | 56 | 1564 |
| 9 | Journal of The American College of Cardiology | 52 | 1.10 | 82 | 2702 |
| 10 | PLOS One | 52 | 1.10 | 0 | 964 |
| 11 | Indian Journal of Pharmacology | 43 | 0.90 | 11 | 130 |
| 12 | Circulation | 39 | 0.80 | 51 | 1388 |
| 13 | Journal of Stroke & Cerebrovascular Diseases | 36 | 0.80 | 43 | 313 |
| 14 | American Journal of Cardiology | 33 | 0.70 | 34 | 567 |
| 15 | Annals of Thoracic Surgery | 33 | 0.70 | 13 | 666 |
| 16 | Indian Journal of Ophthalmology | 33 | 0.70 | 5 | 105 |
| 17 | Atherosclerosis | 32 | 0.70 | 35 | 666 |
| 18 | European Heart Journal | 32 | 0.70 | 27 | 551 |
| 19 | Lancet | 31 | 0.70 | 160 | 24289 |
| 20 | Catheterization And Cardiovascular Interventions | 30 | 0.60 | 11 | 468 |
| 21 | Indian Journal of Pediatrics | 30 | 0.60 | 10 | 240 |
| 22 | Gene | 29 | 0.60 | 28 | 324 |
| 23 | Cerebrovascular Diseases | 28 | 0.60 | 16 | 73 |
| 24 | Biomedicine & Pharmacotherapy | 27 | 0.60 | 10 | 370 |
| 25 | Current Science | 25 | 0.50 | 34 | 477 |
| 26 | European Journal of Pharmacology | 24 | 0.50 | 72 | 648 |
| 27 | Life Sciences | 24 | 0.50 | 52 | 552 |
| 28 | Atherosclerosis Supplements | 23 | 0.50 | 0 | 3 |
| 29 | Clinica Chimica Acta | 23 | 0.50 | 56 | 562 |
| 30 | Biomedical Research-India | 22 | 0.50 | 5 | 58 |
|  | Total of 30 Journals | 1447 | 30.80 | 1432 | 45143 |
|  | Total Indian Journal Output | 4698 | 100.00 | 4038 | 120198 |
|  | Share of 30 journals in Indian journal output | 30.80 | 30.80 | 35.46 | 37.56 |

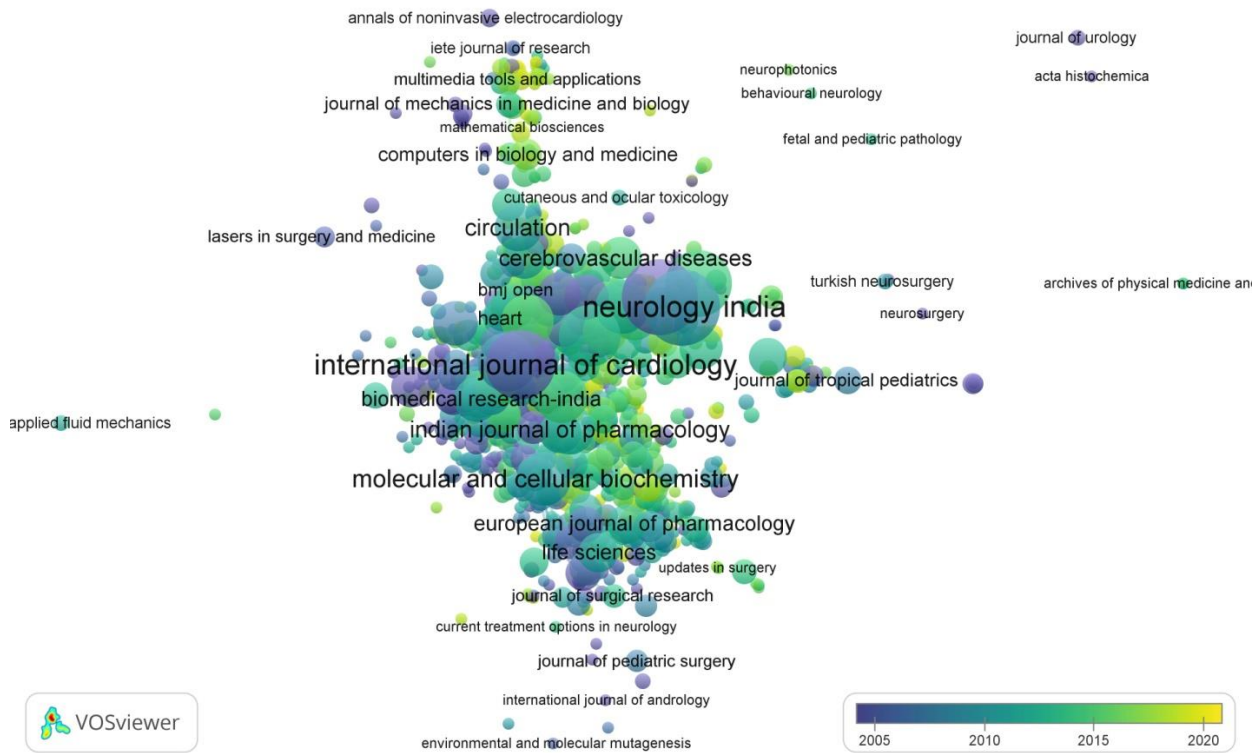

*Cluster of Most Productive Journals in Indian Coronary Artery Disease Research*

**Summary and Conclusion**

1168 Indian publications in coronary artery disease research as indexed in Web of Science database, was published during 1990-2019 and they increased from 15 to 390 in the year 1990 to the year 2019, registering 11.47% growth per annum. Their cumulative Indian output increased from 331 to 1136 to 3231, witnessing 13.12% and 11.02% growth respectively from 1990-1999 to 2000-2009 to 2010-2019. India's global publications share in coronary artery disease research was only 1.14% during 1990-2019, witnessing increase from 0.48% to 0.85% to 1.55% from 1990-1999 to 2000-2009 to 2010-2019. The citation impact per paper of Indian publications on coronary artery disease research was averaged to 25.58 citations, however, increasing from 21.46 during 1990-1999 to 30.38 during 2000-2009 and then decreasing from 30.38 during 2000-2009 to 24.32 during 2010-2019.

The share of India's international collaborative publications in coronary artery disease research was 38.88% during 1990-2019, showing increase from 0.89% during 1990-1999 to 6.00% during 2000-2009 and then again increased to 31.99% during 2010-2019. USA in India's international collaborative papers, contributed the largest publications share of 40.45%, followed by England

(14.72%), Canada (14.07%), Peoples Republic of China (10.73%), Australia (10.24%), and Germany (9.80%) during 1990-2019.

Cardiovascular system and cardiology, among main subjects contributed the highest publications share (19.14%), followed by neurosciences and neurology (14.94%), pharmacology and pharmacy (8.51%), general and internal medicine (4.40%), biochemistry and molecular biology (4.22%), research and experimental medicine (3.81%), surgery (3.23%), cell biology (2.96%), endocrinology and metabolism (2.76%), pediatrics (2.12%) and other subjects respectively during 1990-2019.

Among leading organizations and authors participating in India's coronary artery disease research, the top 20 organizations and top 30 authors together contributed 40.17% and 37.29% respectively as their share of Indian publication output and 36.54% and 38.36% respectively as their share of Indian citation output during 1990-2019. The leading organizations in research productivity were: All India Institute of Medical Sciences (AIIMS), New Delhi (496 papers), Postgraduate Institute of Medical Education & Research (PGIMER), Chandigarh (212 papers), Christian Medical College & Hospital, Vellore (121 papers), Sree Chitra Tirunal Institute Medical Science & Technology (105 papers), Nizams Institute of Medical Science (104 papers), Sanjay Gandhi Postgraduate Institute Medical Science (100) during 1990-2019. The leading authors in publication productivity were Singh S (80 papers), Sharma A (78 papers), Kaul S and Mohan V (74 papers each), Kumar P (70 papers), Singh N and Singh RB (68 papers each), Singh M (63 papers), Gupta R (62 papers), Gupta A (61 papers), and Prabhakaran D (59 papers) during 1990-2019.

Among the total journal output of 4698 papers, the top 30 journals publishing Indian papers in coronary artery disease together accounted for 30.80% share of total Indian journal publication output during 1990-2019. Among journals contributing to Indian coronary artery disease research, *Neurology India* was the most productive journal with 143 papers, followed by *International Journal of Cardiology* (139 papers), *Annals of Indian Academy of Neurology* (99 papers), *Journal of the Neurological Sciences* (77 papers), *Indian Journal of Medical Research* (69 papers), *Molecular And Cellular Biochemistry* (56 papers) Stroke (55 papers), *Journal of the American College of Cardiology* and *PLOS One* (52 papers each) etc. during 1990-2019.

Concludes that coronary artery disease research have been a neglected subspecialty in India, both in teaching and research. There is an urgent need to increase the publication output, improve research quality and improve international collaboration. Review coronary artery disease studies in India indicate that this has become an important public health problem in India. CAD is one of the most important causes of mortality and morbidity in the country. With higher patient coming for treatment and shortage of trained cardiologist specialists are some of the challenges that confront coronary artery disease research at the national level. To address the problems with coronary artery disease research in India, Indian government needs to come up with a policy for identification, screening, diagnosis and treatment of coronary artery disease patients, besides curriculum reform in teaching, capacity building, patient education and political support are badly needed. There is an urgent need to promote primordial, primary, and secondary prevention strategies. Primordial strategies such as promotion of smoking/tobacco cessation, physical activity, and healthy dietary habits should prevent risk factors from occurring in the first place. Primary prevention should focus on screening and better control of risk factors (hypertension, hypercholesterolemia, and diabetes) to prevent incidence of overt CAD. Good quality secondary prevention and better management of acute and chronic events will prevent premature mortality and morbidity.